\documentclass[prb,aps,reprint,superscriptaddress,amsmath,amssymb]{revtex4-2}
\usepackage[latin9]{inputenc}
\usepackage{graphicx}
\usepackage[linkcolor=blue,urlcolor=blue,citecolor=blue,colorlinks=true]{hyperref}

\usepackage{soul}
\usepackage{MnSymbol}
\usepackage[usenames]{color}
\usepackage[dvipsnames]{xcolor}
\usepackage{dcolumn}% Align table columns on decimal point
\usepackage{bm}% bold math

\def\BiCO{{Bi$_2$CuO$_4$}} %Bi$_2$CuO$_4$$\mathrm{Bi_2CuO_4}$
\def\Ig{I_{\mathbf{g}}}
\def\If{I_{\mathbf{f}}}
\def\Alc{\alpha_{\perp}}
\def\Ala{\alpha_{\parallel}}
\def\Eg{E_{\mathbf{g}}}
\def\Ef{E_{\mathbf{f}}}
\def\SR{\hat{S}_{\mathbf{R}}}
\def\SRA{\hat{S}_{\mathbf{R}+\boldsymbol\rho_A}}
\def\SRB{\hat{S}_{\mathbf{R}+\boldsymbol\rho_B}}
\def\SRBf{\hat{S}_{\mathbf{R-f}+\boldsymbol\rho_B}}
\def\SRg{\hat{S}_{\mathbf{R+g}}}
\def\SRf{\hat{S}_{\mathbf{R+f}}}
\def\SqA{\hat{S}_{\mathbf{q}A}}
\def\SmqA{\hat{S}_{-\mathbf{q}A}}

\def\SmqB{\hat{S}_{-\mathbf{q}B}}
\def\EP{\epsilon}
\def\Rbf{\mathbf{R}}
\def\Nbf{\mathbf{n}}
\def\Qbf{\mathbf{q}}
\def\Abf{\mathbf{a}}
\def\Bbf{\mathbf{b}}
\def\Bb{\mathfrak{b}}
\def\Mmu{\mathfrak{m}}

\begin{document}
\title{
Green's function theory of magnetism in Bi$_2$CuO$_4$: anisotropic Heisenberg XYZ model 
} 
\author{R.O. Kuzian}
\affiliation{Donostia International Physics Center (DIPC), Paseo Manuel de Lardizabal
4, San Sebasti\'an/Donostia, 20018 Basque Country, Spain}
\affiliation{Frantsevich Institute for Problems of Materials Science National Academy
of Science of Ukraine, 3, str.\ Omeliana Pritsaka, 03142 Kyiv, Ukraine}
\author{E.E.\ Krasovskii}
\affiliation{Donostia International Physics Center (DIPC),
 Paseo Manuel de Lardizabal 4, San Sebasti\'an/Donostia,
  20018 Basque Country, Spain}
\affiliation{Departamento de Pol{\'i}meros y Materiales Avanzados: F{\'i}sica, Qu{\'i}mica y
  Tecnolog{\'i}a, Universidad del Pa{\'i}s Vasco-Euskal Herriko Unibertsitatea, Donostia-San
  Sebasti\'an, 20080 Basque Country, Spain}
\affiliation{IKERBASQUE, Basque Foundation for Science, 48013 Bilbao, Spain}

\begin{abstract}
The Green's function theory of Lymar' and Rudoi 
[Theor. Math. Phys. \textbf{21}, 990 (1974)] is generalized to the case of multiple 
intra- and inter-sublattice magnetic interactions in a collinear spin-half
antiferromagnet and applied to magnetic excitations in \BiCO. The spin Hamiltonian
includes both the out-of-plane and in-plane symmetric anisotropy terms and the 
Zeeman term, which describes the interaction with an external magnetic field $B$ 
applied along the N\'eel vector. 
Within the spin-wave approximation we calculate spin excitation dispersion, 
antiferromagnetic resonance frequencies and the critical field $B_c$ of the spin-flop 
metamagnetic transition. A weak in-plane anisotropy is shown to result in a gap in the
acoustic-like branch of the excitations. The gap nonlinearly depends on the external field
and closes at $B=B_c$. An expression for the N\'eel temperature in the Tyablikov random phase 
approximation at zero field is derived. For the parameters derived from the recent inelastic 
neutron scattering study by Yuan {\it et al}. [Phys. Rev. B \textbf{103}, 134436 (2021)] it 
gives $T_{\rm N}\approx 52$~K.
\end{abstract} 
\date{\today}
\maketitle

\section{Introduction\label{sec:Intro}}

Cuprates are a rich family of ternary copper oxides with a plethora of unusual 
physical properties. The CuO$_4$ plaquette is a building block of various structures, 
such as two-dimensional (2D) corner-shared CuO$_2$ planes of high-$T_c$ superconductors
\cite{Plakida_book,Chen2019,Li2021,Valkov2021,Sobota2021,Bacq-Labreuil2025}, 
one-dimensional (1D) corner-shared \cite{Schlappa2012,Schlappa2018,Chen2021}
and edge-shared \cite{Drechsler2007,Nishimoto15,Johnston2016,Agrapidis2025}
chains or peculiarly arranged isolated plaquettes.
Close to the Fermi energy the electronic structure of cuprates is determined by the Cu $3d$ 
and oxygen $2p$ states of the CuO$_4$ plaquettes. Moreover, the Cu $3d$ states are so
strongly split by the oxygen ligand field that only the $d_{x^2-y^2}$ orbital matters (the axes 
$x$ and $y$ point from the Cu atom to the nearest oxygen atoms). Thus, a mean-field band 
structure of cuprates often contains a single band crossing the Fermi level. 
However, the cuprates with a half-filled band are insulators because of 
strong many-body correlations. Low-energy electron structure of these compounds
is described by effective Hamiltonians, which include both rotationally invariant
Heisenberg exchange interactions between the localized spins of the Cu$^{2+}$ ions 
and anisotropic exchange terms, the single-ion anisotropy terms being prohibited 
for the spin-half systems. Thus, the insulating cuprates represent a physical realization 
of many models of quantum magnetism on lattices of different dimensions and geometries.

\begin{figure}[htb] %%%%%%%%%%%%%%%%%%%%%%%%%%%%%%%%%%%%%%%%%%%%%%%%%%%%%%%%%%%%%%%% 
\includegraphics[width= 0.3\columnwidth]{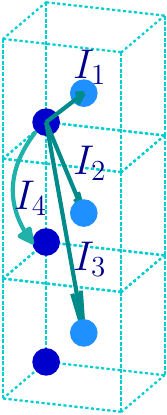}
\caption{\label{f:str} 
The unit cell of \BiCO magnetic lattice contains two Cu$^{2+}$ ions,
having localized spin-half moments aligned in opposite directions 
below N\'eel temperature. Bravais lattice is simple cubic.
Schematically shown are the inter-site exchange interactions
included in the present model following Ref.~\cite{Yuan2021}.
The parameters are: $I_1^x=I(1-\Alc)$, $I_1^y=I(1-\Ala)$,
$I_1^z=I(1+\Ala)$, $I=4.7$~meV $\Alc=0.013$, $\Ala=4.7\cdot 10^{-6}$;
the interactions labeled $I_{2-4}$ are assumed to be isotropic with the parameters
$I_2=1.1$~meV, $I_3=0.5$~meV for inter-sublattice interactions, 
and $I_4=0.36$~meV for a frustrating intra-sublattice intra-chain
interaction .
}
\end{figure} %%%%%%%%%%%%%%%%%%%%%%%%%%%%%%%%%%%%%%%%%%%%%%%%%%%%%%%%%%%%%%%%%%%%%%% 
The \BiCO\ crystal is a special one because, while having the same chemical formula 
and ions valences as the high-$T_c$ compounds $R_2$CuO$_2$ ($R$ stands for La or a rare-earth 
ion), it has completely different structure and properties. In the tetragonal crystal 
structure of Bi$_2$CuO$_4$ the CuO$_4$ plaquettes are stacked along the $c$ axis, the plaquette 
plane being perpendicular to the  
axis~\cite{Boivin,Garcia-Munoz1990}. The neighboring in-chain plaquettes 
are twisted with respect to each other by about 33$^{\circ}$ \cite{Garcia-Munoz1990}. 
The in-chain Cu-Cu distance is much shorter than the 
inter-chain one, see Fig.~\ref{f:str}. That is why \BiCO\ was originally considered as a 
quasi-1D spin-half compound \cite{Sreedhar1988,Attfield1989}. Further studies established
that \BiCO\ is a 3D antiferromagnet with the N\'eel temperature $T_{\mathrm{N}}$
in the range 42--47~K 
\cite{Garcia-Munoz1990,Troc1990,Ong1990,Yamada1991,Konstantinovic1991}. 
Inelastic neutron scattering (INS) studies 
\cite{Ain1993,Roessli1993,Roessli1997,Yuan2021}
unambiguously evidence that the inter-chain superexchange interactions are substantially 
stronger than the intra-chain ones. This statement was also 
supported by the parameter-free density functional theory (DFT) calculation of 
Ref.~\cite{Janson2007}, which has demonstrated that the naive idea of the interaction 
decreasing with distance is erroneous.

Even more dramatic is the story of the determination of the ordered 
moment direction in \BiCO, which seems to be far from finished. The moment 
orientation parallel to the crystallographic direction $\mathbf{c}$ was proposed 
in the neutron diffraction studies of Refs.~\cite{Garcia-Munoz1990,Konstantinovic1991}. 
This model was supported by the optical spectroscopy experiments of 
Refs.~\cite{Konstantinovic1992,Konstantinovic1994,Konstantinovic1996}, but it 
contradicts the magnetization studies of Ref.~\cite{Yamada1991}, which show 
a linear dependence of the magnetization on external magnetic field 
$\mathbf{B}\parallel\mathbf{c}$ and a metamagnetic spin-flop transition 
for $\mathbf{B}\perp\mathbf{c}$. Also according to 
the antiferromagnetic resonance (AFMR) \cite{Ohta1992,Pankrats1994,Pankrats1998,Ohta1998} 
and torque magnetometry measurements \cite{Herak2010} the moment lies in the $\mathbf{ab}$ 
plane, $\Abf\parallel [100]$ and $\Bbf\parallel [010]$. This 
long-standing controversy was resolved by the polarized neutron scattering measurements of 
Ref.~\cite{Zhao2017} that definitely showed that the moment lies in the 
$\mathbf{ab}$ plane. 
However, the exact direction of the moment in the plane is still under question. The 
magnetization measurements reported in Ref.~\cite{Zhao2017} indicate that the moment 
lies either along $\Abf$ or along $\Bbf$. At the same time, the neutron 
scattering in the applied field~\cite{Yuan2021} and recent nonreciprocal directional 
dichroism results~\cite{Kimura2022,Miyamoto2025} are interpreted within the assumption 
that the moment is parallel to either $[110]$ or $[1\bar{1}0]$ direction.

Besides its interest for fundamental physics, the recently discovered magnetically 
induced ferroelectricity and linear magnetoelectric coupling \cite{Zhao2017} opens a way 
to control the direction of the spontaneous magnetic moment by electric field, which makes
\BiCO\ a model material to explore the perspective of higher-capacity memories 
\cite{Kimura2022,Miyamoto2025}.

In this paper we address the magnetism in \BiCO\ with a two-time temperature Green's functions 
(GF) method \cite{Zubarev60,Tyablikov,Zubarev96,Rudoy2011}. We follow the work of 
Lymar' and Rudoi (LR) \cite{Lymar1974}, who considered an anisotropic two-sublattice 
antiferromagnet (XYZ-model) in a longitudinal external magnetic field and included 
only nearest-neighbor non-frustrated inter-sublattice exchange interaction. 
Here we generalize the LR theory by including multiple inter- and intra-sublattice 
interactions. The latter are frustrating if their exchange parameters have antiferromagnetic 
(AFM) sign. The present formalism is an improvement over that of Ref.~\cite{Janson2007}, 
in which the easy-axis XXZ model was assumed. We adopt the model 
proposed in Ref.~\cite{Yuan2021} with four inter-site exchange interactions, see 
Fig.~\ref{f:str}, and show that the spectrum of excitations above a collinear 
AFM ground state consists of two gapped branches. The widths of the gaps 
depend on the longitudinal external magnetic field and are proportional to the AFMR 
frequencies \cite{Ohta1992,Pankrats1994,Pankrats1998,Ohta1998}. 
For a general XYZ model the field dependence is non-linear in contrast to
the easy-axis XXZ model, where the linear dependence on the field makes 
the critical field of the spin-flop transition proportional to the gap 
width at zero field \cite{Feder1968}. We generalize Eq.~(3.7) of the LR 
theory \cite{Lymar1974}, which relates the critical field to the smallest 
of the anisotropy parameters. Thus, we show that the anisotropy parameters 
may be found from AFMR and magnetization 
measurements, which are independent of INS ones.

The paper is organized as follows. In Sec.~\ref{sec:GF}, we introduce the model 
and obtain the retarded GF in the Tyablikov (RPA) approximation 
\cite{Tyab1,Tyablikov,Lymar1974}. The low temperature spectrum in the spin-wave 
approximation is analyzed in Sec.~\ref{sec:LSWT}. In Sec.~\ref{sec:Aniso},  
we find the anisotropy parameters and in Sec.~\ref{sec:TN} calculate the N\'eel 
temperature. The discussion of the results, comparison with the previous works, and
conclusions are in Sec.~\ref{sec:Con}. Appendix~\ref{sec:Adtls} contains 
details of the GF calculation, Appendix~\ref{sec:ALSWT} shows the equivalence 
of the GF approach and linear spin-wave theory (LSWT) at low temperatures.
In Appendix~\ref{sec:SQL} we compare our theory with that of Ref.~\cite{Yuan2021}
in application to a simplified model with $I_2=I_3=I_4=0$.

\section{Green's functions for XYZ spin-Hamiltonian\label{sec:GF}}

Below we calculate the retarded GF in the Tyablikov approximation for the 
XYZ model. We will generalize the consideration of Ref.~\cite{Lymar1974}
by taking into account multiple exchange interactions and 
the specific geometry of \BiCO\ . The model is given by the anisotropic 
Heisenberg Hamiltonian
\begin{align}
\hat{H} & =  \hat{H}_A+\hat{H}_B+\hat{H}_{AB}+\hat{H}_Z, \label{H}\\
\hat{H}_{A(B)} & =\frac{1}{2}\sum_{\Rbf\in A(B)}\sum_{\mathbf{g},\gamma}
\Ig^{\gamma}\SR^{\gamma}\SRg^{\gamma}   \label{eq:HA},\\
\hat{H}_{AB} & =  \sum_{\Rbf\in A}\sum_{\mathbf{f},\gamma}
\If^{\gamma}\SR^{\gamma}\SRf^{\gamma} \label{eq:HAB}, \\
\hat{H}_Z & =-g_z\mu_{\mathrm{B}}B\sum_{\Rbf\in A}(\SR^z+\SRB^z), \label{eq:HZ}
\end{align}
where $\gamma =x$, $y$, or $z$, and $\Rbf$ runs over sites in one sublattice, 
where the copper spins $S=1/2$ are located. Vector $\mathbf{g}$ connects interacting 
sites in the same sublattice and $\mathbf{f}$ between the sublattices. The special 
feature of the magnetic structure of \BiCO\ is that the magnetic unit cell is 
half of the structure unit cell. It contains two sublattices,
$\Rbf_A=\Nbf+\boldsymbol\rho_A$ and $\Rbf_B=\Nbf+\boldsymbol\rho_B$,
where $\Nbf=n_{1}\Abf+n_{2}\Bbf+n_{3}\mathbf{c}/2$ are 
lattice translations 
in terms of the basis vectors $\Abf$, $\Bbf$, and $\mathbf{c}$ of the 
tetragonal lattice. The positions of the two sites are $\boldsymbol\rho_A=0$ and 
$\boldsymbol\rho_B=\Abf/2+\Bbf/2-2z_{\rm Cu}\mathbf{c}$, with 
$z_{\rm Cu}=0.0766$ \cite{Garcia-Munoz1990,Yamada1991}, see Fig.~\ref{f:str}. 
The direction of the spin quantization axis $\hat{z}$ is parallel to the 
N\'eel vector $\mathbf{l}\equiv \langle\hat{\mathbf{S}}_{\Rbf_A}\rangle 
-\langle\hat{\mathbf{S}}_{\Rbf_B}\rangle$, 
where $\langle\hat{\mathbf{S}}_{\Rbf_{A(B)}}\rangle$ is the spontaneous
spin moment proportional to the sublattice magnetization. As mentioned in the 
introduction, its direction is still not established: According to 
Ref.~\cite{Zhao2017}, $\hat{z} \parallel \Abf$, while according to
Refs.~\cite{Yuan2021,Kimura2022,Miyamoto2025}, $\hat{z} \parallel \Abf+\Bbf$. 
Our results are the same for both choices of $\hat{z}$. We then choose 
$\hat{x}=-\hat{c}$ and $\hat{y}=\hat{z}\times\hat{x}$. We follow the model proposed 
in Ref.~\cite{Yuan2021}, which contains a set of four interaction 
bonds. Only the interaction 
along one of the bonds, labeled $I_1$ in Fig.~\ref{f:str}, is anisotropic. 
For $\hat{z} \parallel \Abf+\Bbf$ our frame coincides with the
$(x^{\prime},y^{\prime},z^{\prime})$ frame of Appendix~2 of
Ref.~\cite{Yuan2021} for $\phi = 0$, so that
$I_1^x=I(1-\Alc)$, 
$I_1^y=I(1-\Ala)$, and
$I_1^z=I(1+\Ala)$. 
and the anisotropy parameters, $\Alc>\Ala >0$, have the same meaning as in 
Ref.~\cite{Yuan2021}.
For the interaction parameters
we adopt the values reported in Ref.~\cite{Yuan2021}: the inter-sublattice interaction
parameters are $I=4.7$, $I_2=1.1$, and $I_3=0.5$~meV, and the parameter of 
the frustrating intra-sublattice intra-chain interaction is $I_4=0.36$~meV. 
The anisotropy parameters $\Alc$ and $\Ala$ will be defined in Sec.~\ref{sec:Aniso}.

The Zeeman Hamiltonian $\hat{H}_Z$ describes the interaction with the external 
magnetic field applied along $\hat{z}$. Its parameters are the $g$-factor of the
Cu$^{2+}$ ion in the $\mathbf{ab}$ plane of \BiCO, $g_z\approx 2.04$, and 
the Bohr magneton $\mu_{\mathrm{B}}\approx 5.7884\cdot 10^{-2}$~meV/T. 
Taking into account the weak magnetic field 
$B < B_c \ll k_{\mathrm{B}}T_{\mathrm{N}}/(g_z\mu_{\mathrm{B}})$ 
at a low temperature $T\ll T_{\mathrm{N}}$ will allow us to 
determine in Sec.~\ref{sec:Aniso} the tiny parameter of the magnetic 
anisotropy in the $\mathbf{ab}$ plane $\Ala$ from the value of the 
spin-flop field $B_c$. 
 
Let us introduce the spin-deviation operators 
\begin{align*}
\SR^{\pm} & =\SR^x\pm i\SR^y, \quad %\\
[\SR^{+},\hat{S}_{\mathbf{R^{\prime}}}^{-}] =2\SR^z\delta_{\mathbf{RR^{\prime}}}, \\
 & [\SR^{\pm},\hat{S}_{\mathbf{R^{\prime}}}^{z}]
=\mp \SR^{\pm}\delta_{\mathbf{RR^{\prime}}},
\end{align*}
and rewrite the Hamiltonians (\ref{eq:HA}), (\ref{eq:HAB}) as 
\begin{align}
\hat{H}_{A(B)} & =\frac{1}{2}\sum_{\Rbf\in A(B)}\sum_{\mathbf{g}}
\Bigl[\Ig^z\SR^z\SRg^z \Bigr. \nonumber \\
 &  %\left. 
+\frac{\Ig}{2}\left(\SR^{+}\SRg^{-}
+\SR^{-}\SRg^{+}\right)  \nonumber\\
 &  \Bigl. + \frac{\Eg}{4}\left(\SR^{+}\SRg^{+}
 +\SR^{-}\SRg^{-}\right)\Bigr],
 \label{eq:HApm} \\
\hat{H}_{AB} & =  \sum_{\Rbf\in A}\sum_{\mathbf{f}}
\Bigl[\If^z\SR^z\SRf^z
+\frac{\If}{2}\left(\SR^{+}\SRf^{-}
+\SR^{-}\SRf^{+}\right) \Bigr.+ \nonumber \\
  & \Bigl.
 + \frac{\Ef}{4}\left(\SR^{+}\SRf^{+}
 +\SR^{-}\SRf^{-}\right)\Bigr],
 \label{eq:HABpm} 
\end{align}
where $\Ig=(\Ig^{x}+\Ig^{y})/2$, $\If=(\If^{x}+\If^{y})/2$, 
$\Eg= \Ig^{x}-\Ig^{y}$, and $\Ef= \If^{x}-\If^{y}$.

We introduce the Fourier transform of the spin operators 
\begin{equation}
\hat{S}_{\Qbf s}^{\alpha}=\frac{1}{\sqrt{N}}\sum_{\Nbf}
\mathrm{e}^{i\mathbf{q(n+}\boldsymbol\rho_{s}\mathbf{)}}
\hat{S}_{\mathbf{n+}\boldsymbol\rho_{s}}^{\alpha},
\label{Sq}
\end{equation}
where $s=A$ or $B$ is the sublattice index, and $\alpha=+$, $-$, or $z$; 
$N$ is the total number of the
\emph{unit cells}, which coincides with the number of sites in a sublattice. We 
will calculate the retarded Green's functions for the spin-deviation operators
\begin{align}
 & G_{s_{1}s_{2}}^{\alpha\beta}(\Qbf ,\omega) 
 \equiv \llangle \hat{S}_{-\Qbf s_{1}}^{\alpha}|
 \hat{S}_{\Qbf s_{2}}^{\beta}\rrangle  \nonumber \\
& =
-i\int_{t^{\prime}}^{\infty}\!\!dte^{i\omega(t-t^{\prime})}
\left\langle \left[\hat{S}_{-\Qbf s_{1}}^{\alpha}(t),
\hat{S}_{\Qbf s_{2}}^{\beta}(t^{\prime})\right]\right\rangle , \label{GF}
\end{align}
where $\alpha,\beta=+$ or $-$, $[\hat{X},\hat{Y}]$ 
is a commutator, and $\langle...\rangle$ denotes the thermal average: 
\begin{equation}
\langle\hat{X}\rangle=Q^{-1}\mbox{Tr}\ [\mathrm{e}^{-\beta\hat{H}}\hat{X}],\
 Q=\mbox{Tr}\ \mathrm{e}^{-\beta\hat{H}}. \label{avdef}
\end{equation}
Here Tr denotes 
the trace of an operator and $\beta=(k_{\mathrm{B}}T)^{-1}$ is the 
inverse temperature. The time dependence of an operator $\hat{X}(t)$ is 
given by $\hat{X}(t)=\mathrm{e}^{it\hat{H}}\hat{X}\mathrm{e}^{-it\hat{H}}$.
We denote
\begin{align*}
G_{ 1}(\Qbf ,\omega) & \equiv\llangle \SmqA^{+}|\SqA^{-}\rrangle , \\
G_{2}(\Qbf ,\omega) & \equiv\llangle \SmqB^{-}|\SqA^{-}\rrangle , \\
G_{3}(\Qbf ,\omega) & \equiv \llangle \SmqA^{-}|\SqA^{-}\rrangle ,\\
G_{4}(\Qbf ,\omega) & \equiv\llangle \SmqB^{+}|\SqA^{-}\rrangle .
\end{align*}

The equations of motion for GFs (\ref{GF}) are
\begin{align}
\omega G_{s_{1}s_{2}}^{\alpha\beta}(\Qbf ,\omega) & = 
 \left\langle \left[\hat{S}_{-\Qbf s_{1}}^{\alpha},
 \hat{S}_{\Qbf s_{2}}^{\beta}\right]\right\rangle \nonumber \\
 & +\llangle \left[\hat{S}_{-\Qbf s_{1}}^{\alpha},
 \hat{H}\right]|\hat{S}_{\Qbf s_{2}}^{\beta}\rrangle . 
 \label{eq:Eqm}
\end{align}
The commutators $[\hat{S}_{-\Qbf s_{1}}^{\pm},\hat{H}]$ are given in 
Appendix~\ref{sec:Adtls}. 

LR~\cite{Lymar1974} analyzed several approximate methods to decouple
the infinite set of equations (\ref{eq:Eqm}). Here, we adopt
the two simplest ones: (i) the LSWT
approximation, which is valid at $T=0$ and is used to interpret the INS 
experiments, and (ii) the Tyablikov self-consistent approximation that 
interpolates the GF in the interval between zero and the N\'eel temperature.

After the Tyablikov decoupling \cite{Tyab1,Tyablikov,Lymar1974} of the higher order GF 
\begin{equation}
\llangle \SR^{+}\hat{S}_{\Rbf^{\prime}}^{z}|
\SqA^{-}\rrangle 
\simeq\left\langle \hat{S}_{\Rbf^{\prime}}^{z}\right\rangle 
\llangle \SR^{+}|
\SqA^{-}\rrangle \label{eq:RPA}
\end{equation}
the equations of motion (\ref{eq:Eqm}) may be written 
in a matrix form
\begin{equation}
\begin{pmatrix}
\EP - \EP_A & -\Mmu_3     &  -\Mmu_1     & -\Mmu_2 \\
-\Mmu_3^{*} & \EP - \EP_B &  -\Mmu_2^{*} & -\Mmu_1 \\   
\Mmu_1      &  \Mmu_2     & \EP + \EP_A  & \Mmu_3 \\
\Mmu_2^{*}  & \Mmu_1      & \Mmu_3^{*}   &\EP + \EP_B 
\end{pmatrix}
\begin{pmatrix}
G_{1}\\ G_{2}\\ G_{3}\\ G_{4}
\end{pmatrix}
=
\begin{pmatrix}
1 \\ 0 \\ 0\\ 0
\end{pmatrix}, \label{eq:EqmBi}
\end{equation}
or, briefly,
\begin{equation}
\mathbf{MG}=\mathbf{K}.\label{eq:EqmBiS}
\end{equation}
Here $\EP \equiv \omega /(2m)$ and 
$m \equiv m_A=\left\langle \hat{S}_{\Rbf_A}^{z}\right\rangle$
is proportional to the magnetization of the sublattice A. 
The LSWT for $S=1/2$ corresponds to a non-self-consistent theory 
with $m=S=1/2$. At low temperatures, $T\ll T_{\mathrm{N}}$, for the 
external field smaller than the spin-flop field, $B < B_c$, the magnetization of a 
monodomain sample is zero. Consequently, the magnetization of the 
sublattice B is opposite and has the same magnitude
$m_B=\left\langle \hat{S}_{\Rbf_B}^{z}\right\rangle=-m_A$. Then 
the other elements of the matrix $\mathbf{M}$ are
\begin{align}
\EP_A & \equiv \Mmu_0 + \Bb,\quad \EP_B \equiv \Mmu_0 - \Bb, \\
\Bb & \equiv g_z\mu_{\mathrm{B}}B/(2m), \\
\Mmu_0 & \equiv \frac{1}{2}\left[\sum_{\mathbf{g}}
(\Ig\mathrm{e}^{i\mathbf{qg}}-\Ig^z)+\sum_{\mathbf{f}}\If^z\right], \\
\Mmu_1 & \equiv  \frac{1}{4}\sum_{\mathbf{g}}\Eg\mathrm{e}^{i\mathbf{qg}},\\
\Mmu_2 & \equiv  \frac{1}{2}\sum_{\mathbf{f}}\If\mathrm{e}^{i\mathbf{qf}},\\
\Mmu_3 & \equiv  \frac{1}{4}\sum_{\mathbf{f}}\Ef\mathrm{e}^{i\mathbf{qf}}.
\end{align}

Solving the system (\ref{eq:EqmBi}) by Cramer's rule, we obtain
\begin{equation}
G_{1}(\Qbf ,\EP) = \frac{Z(\EP)}{\det \mathbf{M}}, \label{eq:G1}
\end{equation}
where $Z(\EP) \equiv \det \mathbf{M^{\prime}}$, with
\begin{equation}
\mathbf{M^{\prime}} \equiv  
\begin{pmatrix}
\EP - \EP_B  & -\Mmu_2^{*}  & -\Mmu_1 \\ 
\Mmu_2       &  \EP + \EP_A & \Mmu_3  \\ 
\Mmu_1       & \Mmu_3^{*}   & \EP + \EP_B 
\end{pmatrix}.
\end{equation}
In the model of Fig.~\ref{f:str} the matrix elements are expressed in terms of the 
following parameters:
\begin{align}
\Mmu_0 & = I_4[\cos(\pi L)-1]+2(I_1^z+I_2+I_3) \label{eq:mu0}\\
\Mmu_1 & =0, \label{eq:mu1}\\
\Mmu_2 & = 2\cos(\pi H)\cos(\pi K)\mathrm{e}^{4i\pi z_{\mathrm{Cu}}L}
  (I_1+ \nonumber \\
  & +I_2\mathrm{e}^{-i\pi L}+I_3\mathrm{e}^{-2i\pi L}),\label{eq:mu2} \\
\Mmu_3 & =  \cos(\pi H)\cos(\pi K)\mathrm{e}^{4i\pi z_{\mathrm{Cu}}L}(I_1^x-I_1^y).
\label{eq:mu3}
\end{align} 
The Cartesian components of the quasimomentum $\Qbf $ are given in the 
reciprocal-lattice units 
of the \emph{structural} unit cell, i.e., $\mathbf{qa}\equiv 2\pi H$, 
$\mathbf{qb}\equiv 2\pi K$, and $\mathbf{qc}\equiv 2\pi L$. 
We recall that in \BiCO\ the structural unit cell is twice as large as the magnetic one.

The determinants in Eq.~\ref{eq:G1} are
\begin{align}
& \det \mathbf{M} = \EP^4-\EP^2(\EP_A^2+\EP_B^2-2|\Mmu_2|^2+2|\Mmu_3|^2) & \nonumber \\
 & +\EP_A^2\EP_B^2-2\EP_A\EP_B(|\Mmu_2|^2+|\Mmu_3|^2)+|\Mmu_2^2-\Mmu_3^2|^2 \label{eq:detM} \\
 & = (\EP^2-\EP_{+}^2)(\EP^2-\EP_{-}^2)  \\
& \det \mathbf{M^{\prime}} =  (\EP^2-\EP_A^2+|\Mmu_2|^2)(\EP + \EP_A) 
%\nonumber \\ & 
-|\Mmu_3|^2 (\EP - \EP_A). \label{eq:detMp}
\end{align}
 
The roots $\pm \EP_{\pm}$ of the equation $\det \mathbf{M} =0$
determine the positions of the poles of the Green function.
Positive roots give the energies of spin excitations,
$\omega_{\pm} =2m\EP_{\pm}$. 
The residues at the poles $Z(\pm \EP_{\pm})$ are important for the 
calculation of the N\'eel temperature.

\section{Excitation spectrum in the linear spin-wave 
approximation \label{sec:LSWT}} 

The linear spin-wave approximation 
is valid at low temperatures, $T\ll T_{\mathrm{N}}$ 
\cite{Bloch1930,Slater1930,Holstein1940,BogoliubovStaty,Oguchi60,Oguchi1971,Toth2015},
see also chapter IV of Ref.~\cite{Tyablikov} and Appendix~\ref{sec:ALSWT}.
It corresponds to $m=S=1/2$, so the excitation 
energies coincide with $\EP_{\pm}$, 
the positive roots of the biquadratic equation~(\ref{eq:detM}): 
\begin{align} 
\EP_{\pm} & =  \sqrt{E_m^2 \pm 2R^2},  \label{eq:E12}\\
E_m^2 & \equiv \frac{\EP_A^2+\EP_B^2}{2}-|\Mmu_2|^2+|\Mmu_3|^2  \nonumber \\
 & = \Bb^2+E_{\mathrm{ea}}^2+|\Mmu_3|^2,  \label{eq:Em}\\
R^2 & = \sqrt{(E_{\mathrm{ea}}\Bb )^2+(\Mmu_0|\Mmu_3|)^2+
 (\Mmu_2\Mmu_3^*-\Mmu_2^*\Mmu_3)^2/4}, \label{eq:R2} \\
E_{\mathrm{ea}}^2 & \equiv \Mmu_0^2-|\Mmu_2|^2. \label{eq:Eea2}
\end{align}
The subscript ``ea'' abbreviates ``easy-axis'' because $E_{\mathrm{ea}}$
is the excitation dispersion for the easy-axis XXZ model, where $I^z>I^y=I^x$, 
or, in other words, $\Ala =\Alc$. Then $\Mmu_3=0$, and the energy dependence on 
the external field becomes linear, so Eq.~(\ref{eq:E12})
acquires the form \cite{Oguchi60,Feder1968}
\begin{equation}
\EP_{\pm,\mathrm{ea}} = E_{\mathrm{ea}}\pm \Bb
=E_{\mathrm{ea}}\pm g_z\mu_{\mathrm{B}}B. \label{eq:epsXXZ}
\end{equation}
In the easy-axis model the two branches are degenerate for $B=0$.

Eq.~(\ref{eq:E12}) shows that in the XYZ model, for $\Ala \neq\Alc$, the spin-wave 
energies depend non-linearly on the external field $B$.

\begin{figure}[htb] %%%%%%%%%%%%%%%%%%%%%%%%%%%%%%%%%%%%%%%%%%%%%%%%%%%%%%%%%%%%%%%% 
\includegraphics[width= 0.45\columnwidth]{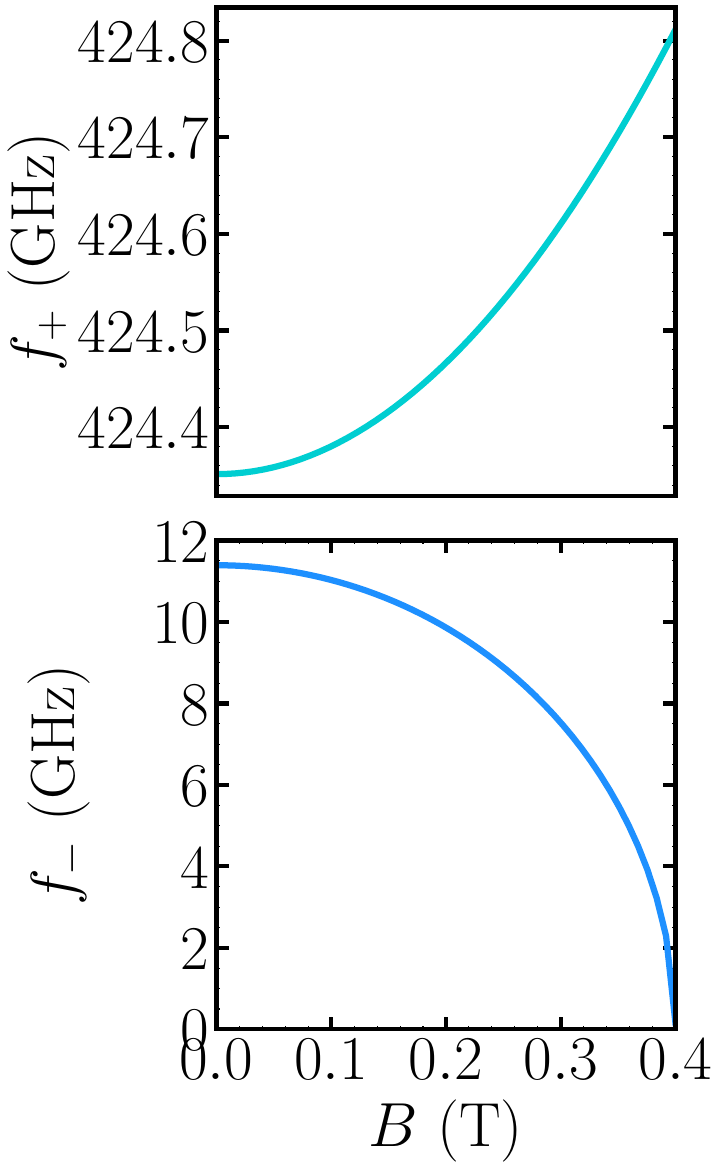}
\caption{\label{f:afmr}
AFMR frequencies  dependence on an external field smaller than the spin-flop 
field, $B < B_c$ 
}
\end{figure} %%%%%%%%%%%%%%%%%%%%%%%%%%%%%%%%%%%%%%%%%%%%%%%%%%%%%%%%%%%%%%%%%%%%%%% 
The gaps at $\Qbf =0$, i.e., at the Brillouin zone (BZ) center $\Gamma$,
determine the two AFMR frequencies for an external field smaller than the spin-flop 
field, $B < B_c$, see Fig.~\ref{f:afmr}:
\begin{align}
\Delta_{\pm} & =2\pi \hbar f_{\pm}, \label{eq:hfpm}\\
\Delta^2_{\pm} & = \Bb^2+ E_{\Gamma,\mathrm{ea}}^2+\Mmu_{3\Gamma}^2 \nonumber \\
 & \pm 2\sqrt{(E_{\Gamma,\mathrm{ea}}\Bb )^2
 +\Mmu_{0\Gamma}^2\Mmu_{3\Gamma}^2} \label{eq:Db}\\
E_{\Gamma,\mathrm{ea}}^2 & \equiv \Mmu_{0\Gamma}^2-|\Mmu_{2\Gamma}|^2,
\end{align}
where the extra subscript ``$\Gamma$'' indicates that 
$\Qbf =0$, i.e., 
\begin{align}
\Mmu_{0\Gamma} & \equiv 2(I_1^z+I_2+I_3) \label{eq:mu00} \\
\Mmu_{2\Gamma} & \equiv 2(I_1+I_2+I_3), \label{eq:mu20} \\
\Mmu_{3\Gamma} & \equiv (I_1^x-I_1^y).\label{eq:mu30} 
\end{align}

\subsection{Dispersion near the BZ center at zero field\label{ssec:Gamma}}

In the close vicinity of
the BZ center, $|H|$, $|K|$, and $|L| \ll 1/2$ we expand the expressions
(\ref{eq:mu0})--(\ref{eq:mu3}), and substitute the 
approximate formulas (\ref{eq:mu0a})--(\ref{eq:mu3a}) derived in Appendix~\ref{ssec:AGamma}
into Eq.~(\ref{eq:E12}). Then the dispersion  
at $\Gamma$ reads
\begin{align}
\EP_{\pm} & \approx \sqrt{\Delta^2_{\pm} + v_{H,\pm}^2Q_{\parallel}^2+v_{L,\pm}^2L^2},
\label{eq:E12a} \\
Q_{\parallel} & =\sqrt{H^2+K^2} \nonumber
\end{align}
where the explicit expressions for the coefficients $v_{H,\pm}$, $v_{L,\pm}$ 
are given in Appendix~\ref{ssec:AGamma}, see Eqs.~(\ref{eq:aHpm}) and 
(\ref{eq:aLpm}).

In the general case, when all the exchange parameters are different, 
$I_i^z > I_i^y \neq I_i^x$, \emph{both} branches of the XYZ model have gaps at zero field. 
In the present model they are
\begin{align}
\Delta_{+} &= 2\sqrt{[I_1^y+I_1^z+2(I_2+I_3)](I_1^z-I_1^x)} \nonumber \\
 & = 2\sqrt{2I(\Alc+\Ala)(I+I_2+I_3)}, \label{eq:D0p}\\
\Delta_{-} &= 2\sqrt{[I_1^x+I_1^z+2(I_2+I_3)](I_1^z-I_1^y)} \nonumber \\
 &= 2\sqrt{2I\Ala [I(2-\Alc+\Ala)+2(I_2+I_3)]}, \label{eq:D0m}
\end{align}
where we have taken into account that $I_1^z > I_1^y > I_1^x >0$, 
i.e., $|I_1^x-I_1^y|= (I_1^y-I_1^x)$.

In the isotropic limit $I_1^z=I_1^x=I_1^y$ (XXX-model, $\Ala = \Alc =0$) both 
branches become gapless. They have acoustic-like dispersions along the high
symmetry directions near $\Gamma$ %the BZ center 
\begin{align}
\EP_{+}^{\mathrm{iso}}(H,K,0) & = \EP_{-}^{\mathrm{iso}}(H,K,0)
 \approx v_H^{\mathrm{iso}}Q_{\parallel}, \\
\EP_{+}^{\mathrm{iso}}(0,0,L) & = \EP_{-}^{\mathrm{iso}}(0,0,L) 
 \approx v_L^{\mathrm{iso}}L.
\end{align}

In the easy-plane case $I_1^z=I_1^y$ (XXZ model, $\Ala =0$, $\Alc > 0$), 
one of the branches is gapless and has an acoustic-like dispersion near 
$\Gamma$,
\begin{align}
\EP_{-}^{\mathrm{ep}}(H,K,0) & \approx v_{H,-}^{\mathrm{ep}}Q_{\parallel},
\label{eq:EPep}\\
\EP_{-}^{\mathrm{ep}}(0,0,L) & \approx v_{L,-}^{\mathrm{ep}}L. 
\end{align}
In contrast, in the easy-axis 
case $I_1^x=I_1^y$ (XXZ model, $\Ala = \Alc > 0$)
the two branches coincide and have a gap
\begin{equation} 
\Delta_{+}^{\mathrm{ea}}=\Delta_{-}^{\mathrm{ea}}=
 4\sqrt{I\Ala (I+I_2+I_3)}. \label{eq:Dea}
\end{equation}

\section{Anisotropy parameters\label{sec:Aniso}}

\begin{figure}[b] %%%%%%%%%%%%%%%%%%%%%%%%%%%%%%%%%%%%%%%%%%%%%%%%%%%%%%%%%%%%%%%
\includegraphics[width=\columnwidth]{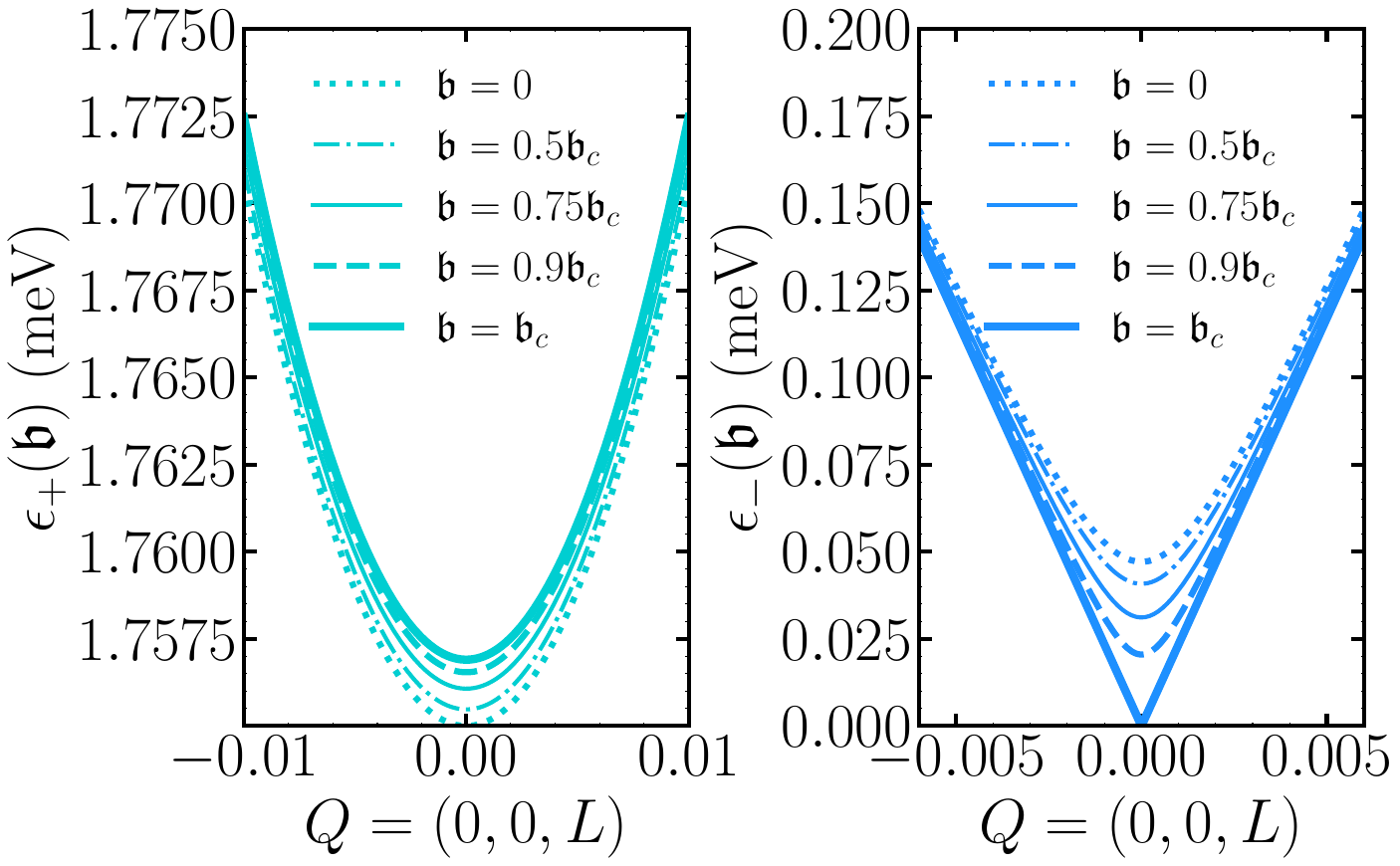} 
\caption{\label{f:wb} Excitation dispersions near the Brillouin zone
center at different external field field $\Bb=g_z\mu_{\mathrm{B}}B$. \textbf{Left:}
the branch $\EP_{+}$ of the spectrum goes up with the increase of the field.
\textbf{Right:} the branch $\EP_{-}$ goes down. The gap in $\EP_{-}$   
vanishes at the critical value of the spin-flop transition 
given by Eq.~(\ref{eq:Bc}). The model parameters are indicated in 
Fig.~\ref{f:str} caption.
}
\end{figure} %%%%%%%%%%%%%%%%%%%%%%%%%%%%%%%%%%%%%%%%%%%%%%%%%%%%%%%%%%%%%%%%%%%%%%% 1

The anisotropy parameters may be found from the knowledge of the two AFMR frequencies 
at an external field smaller than the spin-flop field, 
$B < B_c$. In Ref.~\cite{Ohta1998} the zero field resonance frequency 
was reported to be $f_{+}\approx 390$~GHz. It corresponds to 
a large gap of $\Delta_{+} \approx 1.6$~meV, which is
close to the value $\Delta_{+} \approx 1.75$~meV 
measured in the INS experiment \cite{Yuan2021}. However, the determination 
of the second, low resonance frequency $f_{-}\sim 10$~GHz
is hindered by a large linewidth \cite{Ohta1992,Pankrats1994,Pankrats1998}.
Because of this we will use another way to find the small 
in-plane anisotropy parameter $\Ala$.

When the external field is applied along the direction of the ordered moment, 
one of the gaps, Eq.(\ref{eq:Db}), shrinks
and the other one widens, see Fig.~\ref{f:wb}. 
The narrow gap $\Delta_{-}(\Bb)$ vanishes at a critical value 
\begin{align}
\Bb_c & =g_z\mu_{\mathrm{B}}B_c = \sqrt{\Mmu_{0\Gamma}^2-(|\Mmu_{2\Gamma}|
+|\Mmu_{3\Gamma}|)^2} \nonumber \\
 & =4\sqrt{\Ala I(I+I_2+I_3)}.
\label{eq:Bc}
\end{align}

Thus, substituting the value $B_c \approx 0.4$~T 
\cite{Yamada1991,Zhao2017,Yuan2021} into Eq. (\ref{eq:Bc}), 
we find the in-plane anisotropy
\begin{equation}
\Ala = \frac{(g_z\mu_{\mathrm{B}}B_c)^2}{16I(I+I_2+I_3)}
\approx 4.7\cdot 10^{-6}.
\end{equation}

Using the value of $\Delta_{+}$ obtained from Eq.~(\ref{eq:D0p}) 
we find the out-of-plane anisotropy parameter
\begin{equation}
\Alc \approx \frac{\Delta_{+}^2}{8I(I+I_2+I_3)}-\Ala \approx 0.013,
\end{equation}
which is very close the value found in Ref.~\cite{Yuan2021}.

The higher AFMR frequency is
$f_{+}=\Delta_{+}/h \approx 424$~GHz, which is in good agreement with
the experimental value 390~GHz \cite{Ohta1998}. The 
lower one, $f_{-}=\Delta_{-}/h \approx 11$~GHz, is in accord with
the estimate of Ref.~\cite{Ohta1992}, see Fig.~12 of that paper.

\section{Calculation of N\'eel temperature\label{sec:TN}}

Linear spin-wave approximation is not applicable at the temperatures 
comparable with the ordering temperature $T_{\mathrm{N}}$. The Tyablikov 
self-consistent approximation, Eq.~(\ref{eq:RPA}),
interpolates the GF in the interval between zero and the N\'eel temperature.
The expression (\ref{eq:G1}) for the GF can be recast as a 
sum of four poles
\begin{align}
G_1(\Qbf ,\EP) & =  \frac{Z(\EP )}{(\EP^2-\EP_{+}^2)(\EP^2-\EP_{-}^2)}= \nonumber \\
 & =   \frac{Z(\EP )}{2R^2} 
\left[\frac{1}{\EP_{+}} \left( \frac{1}{\EP - \EP_{+}}-\frac{1}{\EP
 + \EP_{+}} \right) \right. \nonumber \\
& \left. - \frac{1}{\EP_{-}} \left(\frac{1}{\EP - \EP_{-}}-\frac{1}{\EP
 + \EP_{-}} \right) \right]. \label{eq:G1e}
\end{align} 

The equation for the  sublattice magnetization is given by
\begin{equation}
\frac{1}{2}-m=\frac{1}{N}\sum_{\Qbf }
\left\langle \SqA^{-}\SmqA^{+} \right\rangle ,  \label{eq:m}
\end{equation}
where the correlation function $\left\langle \SqA^{-}\SmqA^{+}\right\rangle$ 
is related to the GF~\cite{Zubarev60,Tyablikov,Zubarev96,Rudoy2011}
\begin{equation}
\left\langle \SqA^{-}\SmqA^{+}\right\rangle
 =\int_{-\infty}^{+\infty}\frac{d\omega}{\mathrm{e}^{\beta\omega}-1}
 \left(-\frac{1}{\pi}\mathrm{Im}G_1(\Qbf ,\omega +i0)\right)
\label{eq:Cq}
\end{equation}
Equations (\ref{eq:m}) and (\ref{eq:Cq}) close the RPA self-consistency
loop. From Eq.~(\ref{eq:G1e}) we obtain
\begin{align}
 & -\frac{1}{\pi}\mathrm{Im}G_1(\Qbf ,\omega +i0) = \nonumber \\
 & = \frac{mZ(\omega/2m )}{2R^2}
 \left\{\frac{1}{\EP_{+}}\left[\delta(\omega-2m\EP_{+})
 -\delta(\omega+2m\EP_{+}) \right] \right. \nonumber \\
  & \left.-\frac{1}{\EP_{-}}\left[\delta(\omega-2m\EP_{-})
 -\delta(\omega+2m\EP_{-}) \right]
  \right\}.
\end{align} 
Then Eq.~(\ref{eq:m}) becomes
\begin{align}
 & \frac{1}{2}-m = \nonumber \\
 &=\frac{1}{N}\sum_{\Qbf }\frac{m}{R^2}
\left\{\frac{1}{\EP_{+}}\left[
\frac{Z(\EP_{+})}{\mathrm{e}^{2\beta m\EP_{+}}-1}
-\frac{Z(-\EP_{+})}{\mathrm{e}^{-2\beta m\EP_{+}}-1}
\right]
\right. \nonumber \\
 &\left. -\frac{1}{\EP_{-}}\left[
\frac{Z(\EP_{-})}{\mathrm{e}^{2\beta m\EP_{-}}-1}
-\frac{Z(-\EP_{-})}{\mathrm{e}^{-2\beta m\EP_{-}}-1}
\right]\right\}. \label{eq:mT}
\end{align} 
At the N\'eel temperature the sublattice magnetization vanishes, $m \to 0$.
For small $m$  Eq.~(\ref{eq:mT}) reduces to
\begin{equation}
\frac{1}{2}-m = \frac{1}{N}\sum_{\Qbf }
\left[
\frac{Z(\EP_{+})+Z(-\EP_{+})}{2\beta R^2\EP_{+}^2}
-\frac{Z(\EP_{-})+Z(-\EP_{-})}{2\beta R^2\EP_{-}^2}
\right]
\end{equation}
and provides the formula for $T_{\mathrm{N}}$:
\begin{align}
 & k_{\mathrm{B}}T_{\mathrm{N}}= 1/F, \label{eq:TN}\\
& F =  \frac{1}{N}\sum_{\Qbf }\frac{1}{R^2}
\left[
\frac{Z(\EP_{+})+Z(-\EP_{+})}{\EP_{+}^2}
-\frac{Z(\EP_{-})+Z(-\EP_{-})}{\EP_{-}^2}
\right] \nonumber \\
& = \frac{1}{N}\sum_{\Qbf }\left[
\frac{\EP_A}{\EP_{+}^2}\left(1+\frac{|\Mmu_3|^2}{R^2}\right)
+\frac{\EP_A}{\EP_{-}^2}\left(1-\frac{|\Mmu_3|^2}{R^2}\right)
\right].
\label{eq:F} 
\end{align} 

As usual, in the thermodynamic limit $N\to \infty$ we pass from summation
to integration over the magnetic BZ. Approximate expressions (\ref{eq:E12a})
may be used for analytic integration near the BZ center, where small values 
of $\EP_{-}$ are encountered in the denominator of the 
integrand. The calculation gives $T_{\mathrm{N}}\approx 52$~K, which is slightly
larger than the experimental value $T_{\mathrm{N}}^\mathrm{ exp} \sim 42$--47~K.

\section{Discussion and conclusion\label{sec:Con}}

\begin{figure}[htb] %%%%%%%%%%%%%%%%%%%%%%%%%%%%%%%%%%%%%%%%%%%%%%%%%%%%%%%%%%%%%%%% 
\includegraphics[width= \columnwidth]{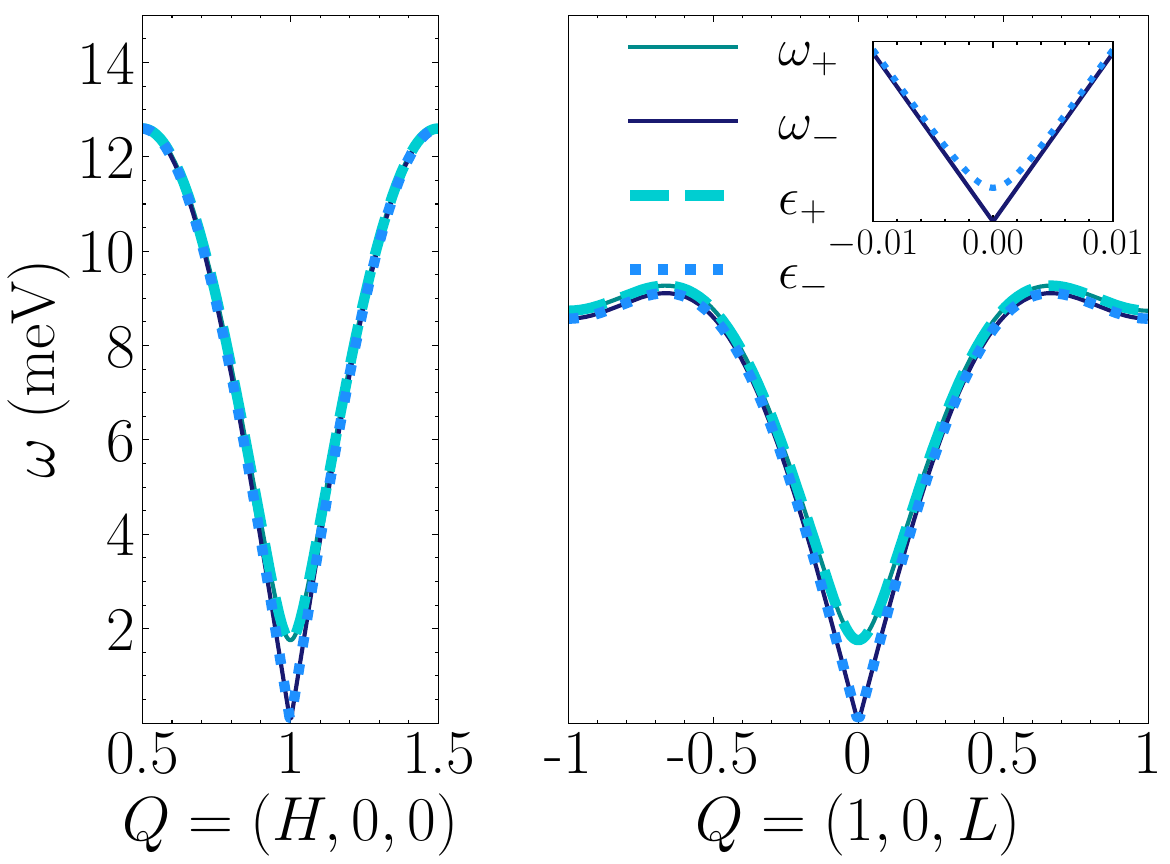}
\caption{\label{f:Y} Comparison of spin-wave dispersion $\EP_{\pm}$, 
Eq.~(\ref{eq:E12}), for zero field, $b=0$,  (dashed and dotted lines) with 
the dispersion $\omega_{\pm}$ given by Eq.~(4) of Ref.~\cite{Yuan2021}
(solid lines). 
The inset shows the vicinity of the Brillouin zone
center $\Gamma$, where the acoustic-like branch $\EP_{-}$
has a gap caused by the in-plane anisotropy, see Eq.~(\ref{eq:E12a}).
The parameter set is the same as in Fig.~\ref{f:wb}
}
\end{figure} %%%%%%%%%%%%%%%%%%%%%%%%%%%%%%%%%%%%%%%%%%%%%%%%%%%%%%%%%%%%%%%%%%%%%%% 
Figure~\ref{f:Y} shows that within the experimental accuracy of $\sim 0.06$~meV 
the dispersion $\EP_{\pm}(\Qbf)$, Eq.~(\ref{eq:E12}), 
coincides with that found in the INS experiments and fitted by $\omega_{\pm}(\Qbf)$ 
given by Eq.~(4) of Ref.~\cite{Yuan2021}. A tiny deviation from $\omega_{\pm}$ may be 
seen in the vicinity of $\Gamma$ because of 
the gap in the acoustic-like branch $\EP_{-}$, see the inset in Fig.~\ref{f:Y}. The gap 
$\Delta_{-} \approx 0.047$~meV is caused by the in-plane anisotropy characterized 
by the parameter $\Ala\approx 4.7\cdot 10^{-6}$. 
In Appendix~\ref{sec:ALSWT}, we demonstrate that within the LSWT the GF approach is
equivalent to the traditional one, which is based on the approximation of the
replacement of the spin-deviation operators by the bosonic ones, see
Refs.~\cite{Bloch1930,Slater1930,Holstein1940,BogoliubovStaty,Oguchi60,Oguchi1971,Toth2015}
and chapter IV of Ref.~\cite{Tyablikov} and references therein. 
Both methods yield the same spectrum $\EP_{\pm}(\Qbf)$, Eq.~(\ref{eq:E12}).
\begin{figure}[htb] %%%%%%%%%%%%%%%%%%%%%%%%%%%%%%%%%%%%%%%%%%%%%%%%%%%%%%%%%%%%%%%% 
\includegraphics[width= 0.7\columnwidth]{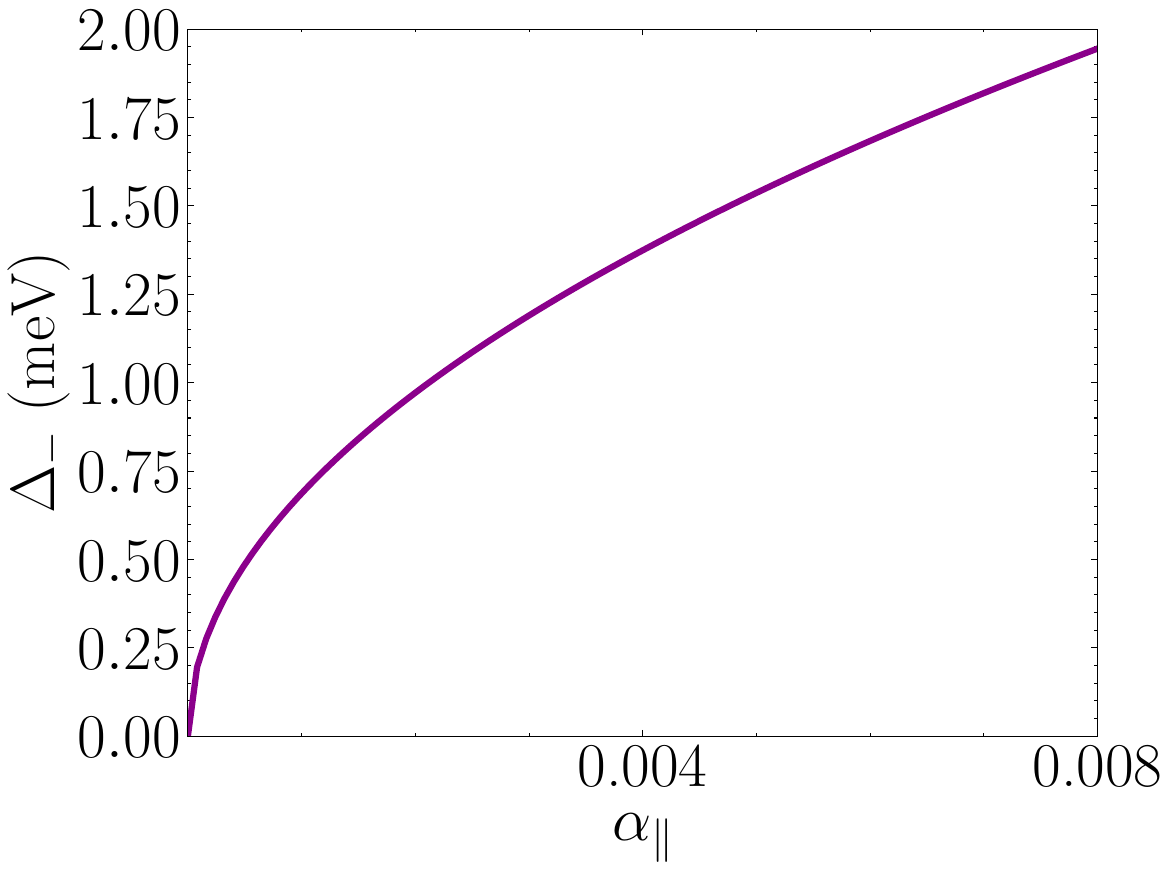}
\caption{\label{f:Delta} The gap in the acoustic-like branch 
$\Delta_{-}=\EP_{-}(0)$ given by Eq.~(\ref{eq:D0m}) on in-plane anisotropy
parameter $\Ala=(I_1^z-I_1^y)/(I_1^z+I_1^y)$.
Other parameters are the same as in Figs.~\ref{f:wb},~\ref{f:Y}.}
\end{figure} %%%%%%%%%%%%%%%%%%%%%%%%%%%%%%%%%%%%%%%%%%%%%%%%%%%%%%%%%%%%%%%%%%%%%%% 

The authors of Ref.~\cite{Yuan2021} 
argue that the tetragonal symmetry of \BiCO\ demands that the acoustic-like branch of 
the dispersion be gapless. However, this statement contradicts their
observation  
that the anisotropic XYZ model is allowed by symmetry in this 
compound. Indeed, the acoustic-like branch $\omega_{-}$ given by Eq.~(4) of 
Ref.~\cite{Yuan2021} remains gapless at \emph{any} 
value of the in-plane anisotropy, including the case $\Ala=\Alc > 0$. However, this 
case corresponds to an easy-axis XXZ model $I_1^z > I_1^x=I_1^y$, which evidently has gaps 
at $B=0$ in the two degenerate branches \cite{Feder1968}, see 
Eq.~(\ref{eq:Dea}). 
In Appendix~\ref{sec:SQL} we elucidate the 
origin of the discrepancy between our 
result for the spin-wave spectrum and that of Ref.~\cite{Yuan2021}.

The gap $\Delta_{-}=\EP_{-}(0)$ given by Eq.~(\ref{eq:D0m})
has a strong square root dependence on $\Ala$, as shown in Fig.~\ref{f:Delta}.

The spin-flop transition in \BiCO\ in the field applied in the $\mathbf{ab}$
plane indicates that the gap in the acoustic-like branch closes
at the critical field $B_c\approx 0.4$~T, cf. Fig.~\ref{f:wb}. The estimate of 
the critical field in Ref.~\cite{Yuan2021} is based on the 
relation $g\mu_{\mathrm{B}}B_{c,\mathrm{Y}}\cos\theta\sim\Lambda$. 
For $\Lambda =0.014\ \mu\mathrm{eV}$ \cite{Yuan2021} and $\cos \theta\sim 1$ it yields 
$B_{c,\mathrm{Y}}\approx 1.2\cdot 10^{-4}\; \mathrm{T} \ll B_c$.
Thus, our description of the magnetism in \BiCO\ is fully 
consistent and is supported by the experimental data.

In Ref.~\cite{Janson2007}, the N\'eel temperature was calculated within the
Tyablikov approximation for the easy-axis XXZ model. As mentioned in the 
introduction, at that time, in 2007, there 
was no certainty regarding the direction of the ordered moments and the 
anisotropy parameters of the \BiCO\ spin-Hamiltonian
nor regarding the parameters of the 
isotropic exchange interactions.
That is why previous $T_{\mathrm{N}}$ calculations were performed for
three different sets of model parameters
collected in rows 3, 5, and 9 of Table I in Ref.~\cite{Janson2007}:
(i) Set~1 was obtained in Ref.~\cite{Janson2007} from DFT calculations;
(ii) Set~2 was a modification of Set~1
that fitted the INS data of Ref.~\cite{Konstantinovic1996};
and (iii) Set~3 was the one originally proposed in Ref.~\cite{Konstantinovic1996}.
These sets were supplemented by the anisotropy parameters that 
reproduced the spin gap $\Delta_{\mbox{\cite{Konstantinovic1996}}}\approx 1.87$~meV
reported in Ref.~\cite{Konstantinovic1996}. The results for the three models
are $T_{\mathrm{N},1}\approx 59$~K and $T_{\mathrm{N},2}\approx T_{\mathrm{N},3}\approx 47$.
The main difference between Set~2 and Set~3 consists in swapping 
(in our notation) $I_1 \rightleftarrows I_3$, 
but it gives almost the 
same spectrum (cf. Fig.~8 of Ref.~\cite{Janson2007}), and, hence, the same 
N\'eel temperature.

The difference of our result 
$T_{\mathrm{N}}\approx 52$~K from that of Ref.~\cite{Janson2007} is 
mainly due to the overall increase of the spectral width. The change of the 
model from the easy-axis to the easy-plane one does not strongly 
affect the result. As a check, we may 
modify our present model and set $\Ala=\Alc=0.013/2$ 
in order to obtain an easy-axis XXZ model with two degenerate branches that 
coincide with $\EP_{+}$. Then the calculation according to Eq.~(\ref{eq:TN}) 
gives $T_{\mathrm{N,ea}}\approx 54$~K. The isotropic version of our model, 
$\Ala=\Alc=0$, yields $T_{\mathrm{N,iso}}\approx 51$~K. As expected, the anisotropy 
suppresses fluctuations and leads to the increase of the ordering 
temperature $T_{\mathrm{N}}$
when the anisotropy terms are added to the isotropic model.

We conclude that the XYZ model consistently describes the spin
excitation spectrum reported in the INS experiments, the metamagnetic
spin-flop transition, and the antiferromagnetic resonance. Still open
remains the reinterpretation of the optical spectroscopy experiments
\cite{Konstantinovic1992,Konstantinovic1994,Konstantinovic1996}, which
were initially interpreted within the assumption that the ordered Cu
spins are parallel to the crystallographic axis $\mathbf{c}$.

\begin{acknowledgments}
Discussions with S.-L. Drechsler and M. D. Kuz'min are gratefully acknowledged.
This work was supported by Spanish MCIU/AEI/10.13039/501100011033,
Projects No.~PID2022-139230NB-I00 and No.~PID2022-138750NB-C22, 
and by the National Academy of Sciences of Ukraine (Project No.~III-6-26).
\end{acknowledgments}

\newpage

\appendix
%dummy comment 

\begin{widetext}

\section{Details of the calculations\label{sec:Adtls}}

\subsection{Commutators\label{ssec:AComm}}
The commutators of operators $\SmqA^{\pm}$ with the Hamiltonian $\hat{H}$ are
\begin{align}
[\SmqA^{\pm},\hat{H}] &= [\SmqA^{\pm},\hat{H}_A]+[\SmqA^{\pm},\hat{H}_{AB}]
+[\SmqA^{+},\hat{H}_Z], \\
[\SmqA^{+},\hat{H}_A] &= \frac{1}{\sqrt{N}}\sum_{\Rbf\in A}
\mathrm{e}^{-i\mathbf{qR}}\sum_{\mathbf{g}}
\left[-\Ig^z\SR^{+}\SRg^z+\Ig\SR^z\SRg^{+}+\frac{\Eg}{2}\SR^z\SRg^{-}\right], 
\label{eq:SqAHA}\\
[\SmqA^{+},\hat{H}_{AB}] &= \frac{1}{\sqrt{N}}\sum_{\Rbf\in A}
\mathrm{e}^{-i\mathbf{qR}}\sum_{\mathbf{f}}
\left[-\If^z\SR^{+}\SRf^z+\If\SR^z\SRf^{+}+\frac{\Ef}{2}\SR^z\SRf^{-}\right], \\
[\SmqA^{-},\hat{H}_A] &= \frac{1}{\sqrt{N}}\sum_{\Rbf\in A}
\mathrm{e}^{-i\mathbf{qR}}\sum_{\mathbf{g}}
\left[\Ig^z\SR^{-}\SRg^z-\Ig\SR^z\SRg^{-}-\frac{\Eg}{2}\SR^z\SRg^{+}\right], 
\label{eq:SqAmHA}\\
[\SmqA^{-},\hat{H}_{AB}] &= \frac{1}{\sqrt{N}}\sum_{\Rbf\in A}
\mathrm{e}^{-i\mathbf{qR}}\sum_{\mathbf{f}}
\left[\If^z\SR^{-}\SRf^z-\If\SR^z\SRf^{-}-\frac{\Ef}{2}\SR^z\SRf^{+}\right],  \\
[\SmqA^{\pm},\hat{H}_Z] &= \pm g_z\mu_{\mathrm{B}}H\SmqA^{\pm}. \label{eq:SqAHZ}
\end{align}
Similarly, the commutators $[\SmqB^{\pm},\hat{H}_B]$, 
$[\SmqB^{\pm},\hat{H}_Z]$ are obtained from 
Eqs.(\ref{eq:SqAHA}), (\ref{eq:SqAmHA}), and (\ref{eq:SqAHZ}) by the 
substitution $\Rbf \to \mathbf{R+\rho}_B$.
The commutators with $\hat{H}_{AB}$ are
\begin{align*}
[\SmqB^{+},\hat{H}_{AB}] &= \frac{1}{\sqrt{N}}\sum_{\Rbf\in A}
\mathrm{e}^{-i\Qbf (\Rbf+\rho_B)}\sum_{\mathbf{f}}
\left[-\If^z\SRBf^z\SRB^{+} %\right. \\ & \left.
+\If\SRBf^{+}\SRB^z+\frac{\Ef}{2}\SRBf^{-}\SRB^z \right], \\
[\SmqB^{-},\hat{H}_{AB}] &= \frac{1}{\sqrt{N}}\sum_{\Rbf\in A}
\mathrm{e}^{-i\Qbf (\Rbf+\rho_B)}\sum_{\mathbf{f}}
\left[\If^z\SRBf^z\SRB^{-} % \right. \\ & \left.
-\If\SRBf^{-}\SRB^z-\frac{\Ef}{2}\SRBf^{+}\SRB^z \right],
\end{align*}

\subsection{Dispersion approximation near the Brillouin zone center\label{ssec:AGamma}}

Near the BZ center $|H|,~|K|,~|L| \ll 1/2$ we may approximately write 
\begin{align}
\Mmu_0 & \approx \Mmu_{0\Gamma}-I_4(\pi L)^2/2,\label{eq:mu0a} \\
\Mmu_2 & \approx \Mmu_{2\Gamma}\left[1-\frac{\pi^2}{2}(H^2+K^2)
-\frac{I_2+4I_3}{\Mmu_{20}}(\pi L)^2\right] 
  -2\pi i(I_2+2I_3)L,\label{eq:mu2a} \\
\Mmu_3 & \approx \Mmu_{3\Gamma}\left[1-\frac{\pi^2}{2}(H^2+K^2)\right]. \label{eq:mu3a}  
\end{align}
Substituting these expressions into Eq.~(\ref{eq:E12}), we obtain for the dispersion
in the center of the BZ, Eq.~(\ref{eq:E12a}). The coefficients are
\begin{align}
v_{H,\pm} & = \pi\sqrt{\Mmu_{2\Gamma}^2-\Mmu_{3\Gamma}^2
\mp |\Mmu_{3\Gamma}|\Mmu_{0\Gamma}} = \\
 & = \pi\sqrt{4(I_1+I_2+I_3)^2-(I_1^x-I_1^y)^2 \pm 2(I_1^x-I_1^y)(I_1^z+I_2+I_3)},
  \label{eq:aHpm}\\
v_{L,\pm} 
 & = \pi\sqrt{-2I_4(I_1^z+I_2+I_3)+4(I_2+4I_3)(I_1+I_2+I_3)-4(I_2+2I_3)^2
 \pm (I_1^x-I_1^y)\left[I_4-\dfrac{(I_2+2I_3)^2}{2(I_1^z+I_2+I_3)}\right]}
  \label{eq:aLpm}
\end{align}
\end{widetext}

\section{Linear spin-wave theory of \BiCO : 
Quadratic form diagonalization\label{sec:ALSWT}}

Here we show that the traditional way to find the LSWT spectrum 
\cite{Bloch1930,Slater1930,Holstein1940,BogoliubovStaty,Oguchi60,Oguchi1971,Toth2015}
(see also chapter IV of Ref.~\cite{Tyablikov}) gives 
the same answer as the GF approach, Eq.~(\ref{eq:E12}). 

LSWT starts from a classical ground state of the spin system. It is 
obtained by substituting spin operators in the Hamiltonian (\ref{H})  
by vectors of the length $S$ ($=1/2$ for our system) and minimizing the 
obtained form with respect to the vectors' directions. For \BiCO\ the 
minimum corresponds to a N\'eel state with two simple cubic sublattices 
A and B. The classical vectors lie in the $\Abf\Bbf$ 
crystallographic plane. 
Here we assume that the vectors are directed along $\Abf+\Bbf$.
LSWT describes the dynamics of small fluctuations
of the spins around their classical direction.
The spin-deviation operators are represented approximately by the bosonic ones 
\cite{Holstein1940}:
\begin{align}
\SRA^{+} & \approx c_{\Rbf_A} =\frac{1}{\sqrt{N}}\sum_{\Qbf}e^{i\Qbf\Rbf_A}c_{\Qbf},
\label{eq:SA2c}
\\ 
\SRA^{-}  & \approx c_{\Rbf_A}^{\dagger} =\frac{1}{\sqrt{N}}\sum_{\Qbf}
e^{-i\Qbf\Rbf_A}c_{\Qbf}^{\dagger},\label{eq:SA2cdagg}\\
\SRA^{z} & \approx \frac{1}{2}-c_{\Rbf_A}^{\dagger}c_{\Rbf_A},\quad
 [c_{\Rbf_A},c_{\Rbf^{\prime}_A}^{\dagger}] =\delta_{\Rbf_A,\Rbf^{\prime}_A}
\label{eq:SAz2c}
\end{align}
And analogously for sublattice B
\begin{align}
\SRB^{+} & \approx d_{\Rbf_B}^{\dagger}=\frac{1}{\sqrt{N}}\sum_{\Qbf}
e^{-i\Qbf\Rbf_B}d_{\Qbf}^{\dagger},\label{eq:SB2d}\\ 
\SRB^{-} & \approx d_{\Rbf_B}=\frac{1}{\sqrt{N}}\sum_{\Qbf}
e^{i\Qbf\Rbf_B}d_{\Qbf},\label{eq:SB2ddagg}\\
\SRB^{z} & \approx -\left(\frac{1}{2}
-d_{\Rbf_B}^{\dagger}d_{\Rbf_B}\right),
\quad
[d_{\Rbf_B},d_{\Rbf^{\prime}_B}^{\dagger}] =\delta_{\Rbf_B,\Rbf^{\prime}_B}.
\label{eq:SBz2d}
\end{align}
Let us note that the expressions (\ref{eq:SA2c})--(\ref{eq:SBz2d}) 
correspond to Eq.~(19)
of Ref.~\cite{Toth2015} with $j=A,B$ and vectors
\begin{equation}
\mathbf{u}_A=
\begin{pmatrix}
1\\ i\\ 0
\end{pmatrix},\;
\mathbf{v}_A=
\begin{pmatrix}
0\\ 0\\ 1
\end{pmatrix},\;
\mathbf{u}_B=
\begin{pmatrix}
1\\ -i\\ 0
\end{pmatrix},\quad
\mathbf{v}_B=
\begin{pmatrix}
0\\ 0\\ -1
\end{pmatrix}.
\end{equation}
Substituting expressions (\ref{eq:SA2c})--(\ref{eq:SBz2d}) into 
Eqs.~(\ref{eq:HApm}) and (\ref{eq:HABpm}) and retaining only the terms
quadratic in bosonic operators we recast the Hamiltonian as 
\begin{align}
\hat{H} &  \approx E_{\mathrm{cl}}+\hat{H}_{\mathrm{SW}}, \quad
E_{\mathrm{cl}}  = \frac{N}{4}\left[\sum_{\mathbf{g}}\Ig^z
-\sum_{\mathbf{f}}\If^z \right], \label{eq:Happr}\\ 
\hat{H}_{\mathrm{SW}} & = \sum_{\Qbf}\Bigl\{\EP_A c_{\Qbf}^{\dagger}c_{\Qbf}
+\EP_B d_{\Qbf}^{\dagger}d_{\Qbf} \Bigr.\nonumber\\
 &  + \frac{\Mmu_1}{2}(c_{\Qbf}^{\dagger}c_{-\Qbf}^{\dagger}
+d_{\Qbf}^{\dagger}d_{-\Qbf}^{\dagger} +c_{\Qbf}c_{-\Qbf} +d_{\Qbf}d_{-\Qbf}) 
\nonumber\\
\Bigl. & +\Mmu_2^{*}d_{\Qbf}^{\dagger}c_{-\Qbf}^{\dagger}
+\Mmu_2c_{-\Qbf}d_{\Qbf} + \Mmu_3^{*}d_{\Qbf}^{\dagger}c_{\Qbf}
+\Mmu_3c_{\Qbf}^{\dagger}d_{\Qbf}
\Bigr\}. \label{eq:HSW}
\end{align}
In this representation the interaction terms like 
$c_{\Rbf_A}^{\dagger}c_{\Rbf_A}c_{\Rbf^{\prime}_A}^{\dagger}c_{\Rbf^{\prime}_A}$
are neglected, which is justified at low temperature when the number of 
excited spin-waves is small, 
$\langle c_{\Rbf_A}^{\dagger}c_{\Rbf_A}+d_{\Rbf_B}^{\dagger}d_{\Rbf_B}\rangle \ll 1$,
or, in other words, 
$1/2 - \langle \SRA^{z}\rangle = 1/2 + \langle \SRB^{z}\rangle \ll 1$.

Thus, the spin-wave Hamiltonian $\hat{H}_{\mathrm{SW}}$ is  quadratic 
in the Bose operators
\begin{equation}
\hat{H}_{\mathrm{SW}}= \frac{1}{2}\sum_{\Qbf}
\bigl(\psi^{\dagger}\mathbf{H}_{\Qbf}\psi - \EP_A-\EP_B\bigr),
\label{eq:Hchin}
\end{equation}
where $\psi^{\dagger} \equiv (c_{\Qbf}^{\dagger},d_{\Qbf}^{\dagger},
c_{-\Qbf},d_{-\Qbf})$, and the matrix $\mathbf{H}_{\Qbf}$ is
\begin{equation}
\mathbf{H}_{\Qbf}=
\begin{pmatrix}
\EP_A & \Mmu_3     &  \Mmu_1     & \Mmu_2 \\
\Mmu_3^{*} & \EP_B &  \Mmu_2^{*} & \Mmu_1 \\   
\Mmu_1      &  \Mmu_2     & \EP_A  & \Mmu_3 \\
\Mmu_2^{*}  & \Mmu_1      & \Mmu_3^{*}   &\EP_B 
\end{pmatrix}. \label{eq:mtrxHq}
\end{equation}
Such a form is diagonalized by a Bogoliubov  
transformation \cite{Bogolyubov1947,Colpa1978,BogoliubovStaty}, which is 
reduced to the diagonalization of the matrix $\mathbf{G}\mathbf{H}_{\Qbf}$, 
where $\mathbf{G}$ is a diagonal $4\times 4$ matrix with the first and 
last 2 entries given by 1 and $-1$, respectively \cite{Toth2015,Yuan2021}. 
It is easy to see that 
$\det(\EP\mathbf{E}-\mathbf{G}\mathbf{H}_{\Qbf}) = \det \mathbf{M}$, where
$\mathbf{E}$ is the identity matrix. Thus, the 
eigenvalues are the roots of the equation $\det \mathbf{M}=0$ that are 
given by Eq.~(\ref{eq:E12}).

\section{Linear spin-wave theory of the spin-\textonehalf{} XYZ model 
on a square lattice \label{sec:SQL}}

% pro chto
As we have discussed in Section~\ref{sec:Con}, the spin-wave spectrum  
given by Eq.~(\ref{eq:E12}) differs from the spectrum by Eq.~(4) of 
Ref.~\cite{Yuan2021}, as shown in the inset of Fig.~\ref{f:Y}. While the difference is  
small for the parameters relevant for \BiCO, it is physically significant. 
The gap in the acoustic-like branch $\Delta_{-}$ given by Eq.~(\ref{eq:D0m}) 
is proportional to the zero-field AFMR frequency $f_{-}$ and is responsible 
for the spin-flop transition at a critical field $B_c$ given by 
Eq.~(\ref{eq:Bc}). 

In Ref.~\cite{Yuan2021} the spectrum has a gapless acoustic branch.
As a consequence, for the explanation of the actual in-plain anisotropy, which 
results in the spin-flop transition, the authors propose a quantum order to 
disorder theory, which is in strong quantitative disagreement with the observed 
critical field $B_c$.  
Here we elucidate the origin of this discrepancy.

First, we see that the matrix  $\mathbf{H}_{\Qbf}$ in Eq.~(\ref{eq:mtrxHq}) 
has the same form as the 
matrix given by Eqs.~(A11)--(A13) in Appendix~2 of Ref.~\cite{Yuan2021} 
(recall that $\Mmu_1=0$).  
It is easy to check that the spin-wave spectrum, Eq.~(\ref{eq:E12}),
for the isotropic model $I_1^x=I_1^y=I_1^z = I$ coincides with that 
given by Eq.~(3) of Ref.~\cite{Yuan2021}. Thus, the difference between 
the spectrum of the XYZ model with 
$I_1^z>I_1^y>I_1^x>0$ and the one given by
Eq.~(4) of Ref.~\cite{Yuan2021}  is caused by  
anisotropic terms. In order to demonstrate its origin,
we consider here a simplified model where all the isotropic interactions in
the Hamiltonian
$\hat{H}$ in Eq.~(\ref{H}) vanish, so that $I_2=I_3=I_4=0$. 
With the nearest-neighbor 
interaction $I_1$ we are left with the model considered in Ref.~\cite{Lymar1974}.
We may set $z_{\mathrm{Cu}}=0$ because the spectrum does not depend
on this parameter despite the fact that 
the matrix elements $\Mmu_2$ and $\Mmu_3$ do depend on $z_{\mathrm{Cu}}$, 
see Eqs.~(\ref{eq:mu2}) and (\ref{eq:mu3}).
We thus arrive at an XYZ model on a square lattice  (SQL)
\cite{Manousakis1991}. In Ref.~\cite{Rosenberg2023} a more general model 
is considered, which, in addition to the magnetic anisotropy, 
is spatially anisotropic.
Below we compare our results with those of Sec.~II of Ref.~\cite{Rosenberg2023}
by setting $v_x=v_y=w_x=w_y=(I_1^x-I_1^y)/2$.

For the SQL at $B=0$ the matrix elements of $\mathbf{H}_{\Qbf}$ are
\begin{align}
\EP_A & = \EP_B=\Mmu_0 = 2I_1^z = 2I(1+\Ala), \label{eq:mu0sq}\\
\Mmu_1 & =0, \label{eq:mu1sq}\\
\Mmu_2 & = (I_1^x+I_1^y)\gamma_{\Qbf}
=I(2-\Ala-\Alc)\gamma_{\Qbf},\label{eq:mu2sq} \\
\Mmu_3 & =  (I_1^x-I_1^y)\gamma_{\Qbf}
=I(\Ala-\Alc)\gamma_{\Qbf}, \label{eq:mu3sq} 
\end{align}
where
\begin{align}
\gamma_{\Qbf} & \equiv \frac{1}{4}\sum_{\mathbf{f}}\mathrm{e}^{i\mathbf{qf}}
=\cos(\pi H)\cos(\pi K) \label{eq:gamq}\\
& =\frac{1}{2}[\cos(q_x\rho_x)+\cos(q_y\rho_y)], \quad % \nonumber\\
\boldsymbol\rho = \frac{\Abf+\Bbf}{2}. \nonumber
\end{align}
Then the model Hamiltonian $\hat{H}_{\mathrm{SW}}$, Eq.~(\ref{eq:HSW}),
coincides with Eq.~(1.5) of Ref.~\cite{Lymar1974} as well as with Eq.~(6)
of Ref.~\cite{Rosenberg2023}.
% zachem gamma_q
We see that the diagonal matrix element $\Mmu_0$ is determined  by the 
largest interaction $I_1^z$, the axis $\hat{z}$ being parallel to the ordered moment
direction. The non-diagonal matrix elements are real and depend on
the wave vector via the function $\gamma_{\Qbf}$. So does 
the spectrum given by Eq.~(\ref{eq:E12}) for the SQL
\begin{equation}
\EP_{\pm}^{\mathrm{SQL}}  = 
2\sqrt{(I_1^z\mp \gamma_{\Qbf}I_1^x)(I_1^z\pm \gamma_{\Qbf}I_1^y)},
\label{eq:EPSL}
\end{equation}
which coincides with Eq.~(3.11) of Ref.~\cite{Lymar1974} and with Eq.~(15)
of Ref.~\cite{Rosenberg2023}. 
Only when $\Ala =0$, so that $I_1^z=I_1^y$, the dispersion has the form
$\EP_{-}^{\mathrm{SQL}}\propto \sqrt{1-\gamma_{\Qbf}}$ and 
is linear near the BZ center 
with the slope proportional to $Q_{\parallel}$, as in Eq.~(\ref{eq:EPep}).
In a general case the dispersion near BZ center is given by Eq.~(\ref{eq:E12a})
with $v_{L,\pm}=0$ and the parameters of the lower branch
\begin{align}
\Delta_{-}^{\mathrm{SQL}} &= 2I\sqrt{2\Ala (2-\Alc + \Ala )}, \\
v_{H,-}^{\mathrm{SQL}} &= \pi I\sqrt{2[2-\Alc-\Ala (3-3\Alc+\Ala )]}. 
\label{eq:vHm}
\end{align}
In the region $\Delta_{-}^{\mathrm{SQL}}/v_{H,-}^{\mathrm{SQL}} 
\sim 2\sqrt{\Ala}/\pi \ll Q_{\parallel} 
\lesssim 0.25$, the dispersion is quasi-linear 
$\EP_{-}^{\mathrm{SQL}}\approx v_{H,-}^{\mathrm{SQL}}Q_{\parallel}$.

The matrix elements given by Eq.~(5) of Ref.~\cite{Yuan2021} 
differ from ours given by Eqs.~(\ref{eq:mu0sq}), (\ref{eq:mu2sq})
and (\ref{eq:mu3sq}).
The correspondence between the interaction parameters $J_{\zeta\zeta}^1$ 
of Ref.~\cite{Yuan2021} and ours is given by the relations 
$J_{zz}^1=I_1^x=I(1-\Alc)$,
$J_{yy}^1=I_1^z=I(1+\Ala)$,  and
$J_{xx}^1=I_1^y=I(1-\Ala)$. 
For comparison we should take the matrix elements in Eq.~(5) of 
Ref.~\cite{Yuan2021} for the case of the ordered moment along 
$\Abf+\Bbf$, i.e., $\phi = 0$. 
Then the diagonal matrix element is 
$C_{\mathbf{\Qbf}}^{\mathrm{SQL}}=J^1_{xx}+J^1_{yy}=2I$.
Despite the statement of the authors of Ref.~\cite{Yuan2021} that the
$z^\prime$-axis is along the direction of the ordered moment,
the diagonal matrix element is not proportional to the largest 
interaction parameter $I(1+\Ala)$ and thus differs
from our Eq.~(\ref{eq:mu0sq}), 
$C_{\mathbf{\Qbf}}^{\mathrm{SQL}} < \Mmu_0$.
Also substantially different from our expressions (\ref{eq:mu2sq}),
(\ref{eq:mu3sq}) is the
wave-vector dependence of the non-diagonal matrix elements 
\begin{align*}
E_{\Qbf}^{\mathrm{SQL}} & = I\mathrm{e}^{-i\pi (H+K)}(\Ala \kappa_{\Qbf} 
-\Alc \gamma_{\Qbf}) \\
F_{\Qbf}^{\mathrm{SQL}} & = I\mathrm{e}^{-i\pi (H+K)}[(2-\Alc)\gamma_{\Qbf}
-\Ala \kappa_{\Qbf}]\\
\kappa_{\Qbf} & \equiv \sin \pi H\sin \pi K.
\end{align*}
Thus, the dependence on the anisotropy parameters of the matrix given 
by Eqs.~(A11)--(A13) in Appendix~2 of Ref.~\cite{Yuan2021} differs from that of our matrix
$\mathbf{H}_{\Qbf}$, see Eq.~(\ref{eq:mtrxHq}), which leads to a different spectrum
\begin{align}
\omega_{\pm}^{\mathrm{SQL}} & 
= \sqrt{(C_{\mathbf{\Qbf}}^{\mathrm{SQL}}\pm |E_{\Qbf}^{\mathrm{SQL}}|)^2
-|F_{\Qbf}^{\mathrm{SQL}}|)^2} \nonumber\\
& = I\{(2\pm |\Ala \kappa_{\Qbf} -\Alc \gamma_{\Qbf}|)^2 \nonumber\\
 & -[(2-\Alc)\gamma_{\Qbf} -\Ala \kappa_{\Qbf}]^2\}^{1/2}.
\label{eq:Yuan}
\end{align}
Near the BZ center the dispersion of the lower branch is linear,
\begin{equation}
\omega_{-}^{\mathrm{SQL}}  \approx vQ_{\parallel}, \quad
v  = \pi I\sqrt{2(2-\Alc)},
\end{equation}
and does not depend on $\Ala$.
However, as mentioned in the main text,
for $\Ala=\Alc > 0$ the model becomes an easy-axis XXZ model 
$I_1^z > I_1^x=I_1^y$ having two degenerate branches with a gap
$\Delta^{\mathrm{ea}}=4I\sqrt{\Ala}$, see Eq.~(\ref{eq:Dea}) and
Ref.~\cite{Feder1968}.

As mentioned earlier, the spectrum of
an XYZ model on an SQL given by equation (\ref{eq:EPSL}) was obtained
previously in Refs.~\cite{Lymar1974,Rosenberg2023}. The present 
derivation is transparent and may be easily reproduced by the reader.

\end{document}